# Reconfigurable Geometric Phase Matching by Multilayered Nonlinear Thin-Film Crystals

**Danielle Ben-Haim[1,2,§,*], Mai Tal[1,2,3,§], Xiaoxi Xu[1,2], Tal Ellenbogen[1,*]**

[§] *Equal contribtuion*

[1] *Department of Physical Electronics, School of Electrical Engineering, Tel-Aviv University, Tel-Aviv 6779801, Israel*

[2] *Center for Light-Matter Interaction, Tel-Aviv University, Tel-Aviv 6779801, Israel*

[3] *Department of Condensed Matter Physics, School of Physics and Astronomy, Tel-Aviv University, Tel-Aviv 6779801, Israel*

[*]*Corresponding author e-mail:* tellenbogen@tauex.tau.ac.il; benhaim.danielle@gmail.com

**Abstract:** Phase matching is essential for efficient energy transfer in nonlinear wave-mixing processes. Traditional methods, such as birefringent and quasi-phase matching, have remained conceptually unchanged since their discovery over 60 years ago, each posing inherent constraints and limitations. Here, we demonstrate the concept of geometric phase matching as a new paradigm for tunable nonlinear wave mixing, based on a multilayered platform of nonlinear thin-film crystals. We leverage this concept to experimentally show reconfigurable and spin-controlled phase matching for second-harmonic generation (SHG), opening new avenues for real-time manipulation of nonlinear interactions in photonic devices. We specifically demonstrate full modulation of SHG from a bilayer structure, nearly perfect and tunable geometric phase matching from an eight-layer structure, and polarization tomography that reveals the evolution of the spin dependent interaction. This approach not only expands the design space for nonlinear optical processes but also paves the way for highly robust, tunable and efficient frequency conversion, for next-generation adaptive nonlinear photonic, quantum photonic and nonlinear optical metamaterial technologies based on thin-film crystals.

Nonlinear optical wave mixing plays a central role in a wide range of advanced scientific and technological developments, including the generation of coherent light throughout the optical spectrum, from extreme ultra-violet up to THz waves[1–3], along with single and entangled photons[4–6], broadband optical frequency combs[7,8], and ultrashort pulses[9]. To promote efficient interaction and coherent buildup of the generated light, it is usually crucial to match the phase of the interacting waves. However, natural material dispersion prevents the phase matching (PM) condition from being satisfied. To overcome this problem, specific PM techniques must be applied. The two most widespread approaches are birefringent PM (BPM), which compensates for refractive index mismatch through material anisotropy, and quasi-PM (QPM), which utilizes periodic modulation of the nonlinear tensor to induce constructive interference of the nonlinear signal. Since their conceptualization, over 60 years ago[10–13], these methods have driven major progress in nonlinear optics, but remain constrained by limited tunability, design flexibility and available material platforms.

Recently, new opportunities for nonlinear wave-mixing were explored in the context of engineered metasurfaces and metamaterials. It was shown that metasurfaces facilitate unprecedented control capabilities over the nonlinear interactions[14–19], however at the cost of limited conversion efficiency due to short interaction lengths. Stacking of metasurfaces into 3D nonlinear metamaterials was also studied, showing improvements in conversion efficiency[20,21]. Yet, due to strong resonant scattering, they show nontrivial collective phenomena[22], reduced transmission, and bands of increased absorption that limit their applicability. Very recently, stacked layers of Van-der-Waals materials were also explored as new means for enhancing frequency conversion, and different types of PM schemes were examined[23–26]. However, their applicability may be hindered by challenging fabrication schemes, limited scalability, and absorption in the visible and near infrared region.

Here we explore multilayered thin-film lithium niobate (TFLN) as a versatile and scalable platform for engineering efficient nonlinear crystals through a generalized PM approach, that shows new adaptive control capabilities over the nonlinear process. This approach is based on geometric phase that arises when light interacts with anisotropic structures, also known as Pancharatnam-Berry phase[27,28]. The circular polarization states experience geometric phase that relies only on the structure's rotational symmetry, contrary to the conventional propagation phase that depends on the material dispersion. In the nonlinear case, specific selection rules govern the allowed harmonic generation processes and dictate the harmonics spin state and corresponding geometric phases[29]. The concept of geometric phase has been instrumental in modern optics, including in the design of ultrathin optical elements such as metasurfaces[30–34] and topological photonic structures[35], and found manifestation in many optical and physical processes[36–38]. Here, by engineering the nonlinear geometric phase in real-time across multiple TFLN layers, we demonstrate broad control over the nonlinear interactions, enabling dynamically tunable, routed, and spin-controlled PM of second-harmonic generation (SHG). This geometric-PM (GPM) technique expands the PM design space beyond the constraints of bulk and periodically polled nonlinear materials, offering an alternative scheme based on the crystal's rotational symmetry, allowing to reach nearly perfect PM over a widely tunable bandwidth, and removing limitations posed by the interaction length and the material dispersion.

**Nonlinear Thin-Film Crystals**

In recent years, the maturation of ion implantation and wafer processing techniques has established nonlinear thin-film crystals as a transformative platform for modern integrated photonics[6,40,41]. As such, Lithium niobate (LiNbO$_3$; LN) has emerged as the dominant thin-film platform for nonlinear optics, due to its strong nonlinear optical response, combined with broad transparency, high damage threshold, and ease of ferroelectric domain engineering[1]. The aforementioned properties, combined with strong optical confinement in thin films, have sparked the development of a broad range of on-chip, active and passive, nonlinear photonic devices[42,43]. Beyond various guided-wave structures, TFLN has recently been harnessed to realize nonlinear metasurfaces[44], paving the way for spatially engineered generation and manipulation of classical and quantum states of light, including demonstrations of nonlinear beam shaping, routing, and holography. All together, these advances position TFLN as a versatile and powerful material for next-generation optical and photonic technologies.

Here we present an alternative strategy for leveraging the advantages and maturity of the TFLN platform by employing a multilayered configuration. Instead of on-chip or metasurface implementation, we use the thin-film characteristics for the construction of a tailored stack of nonlinear thin-films, as an efficient and highly tunable phase-matched nonlinear platform with new control capabilities. This approach is made possible due to the three-fold (C3) rotational symmetry of LN around its z-axis (see Fig. 1a), giving rise to nonlinear geometric phase in a spin-selective SHG process. It was shown, mostly in the framework of nonlinear metasurfaces, that the illumination of C3 symmetric materials with circularly polarized fundamental frequency (FF), leads to the generation of cross-circularly polarized second-harmonic (SH)[14,26,45]. Notably, the generated SH carries a geometric phase of three times the rotation angle around the material's symmetry axis, and changes sign with the spin state (see Fig. 1b). This geometric phase was utilized in nonlinear metasurfaces for a plethora of exciting demonstrations, e.g., nonlinear beam shaping, scalar and vectorial holography, image encoding, and functional THz emitters, just to name a few [14,46–49]. Here, we show that by transitioning to bi- and multi-layer structures, the geometric phase serves as a new means for real-time controlled PM of nonlinear interactions, opening the door for efficient reconfigurable spin-controlled nonlinear optics.

**The Concept of Geometric Phase Matching**

The SHG process is described by a set of coupled-wave-equations[50]. We consider the FF wave propagating along the z direction, in a dispersive material with refractive index n(ω). Under the assumption of non-depleted pump, forward propagating waves, and slowly varying envelope approximation, the coupled equations are reduced to a single equation of the form[50]:

$$\frac{dE^{SH}}{dz} = \frac{i2\omega}{n(2\omega)\,c} d_{eff}(E^{FF})^2 e^{i\Delta k z} \quad (1)$$

where $d_{eff}$ is the effective second-order nonlinear coefficient, $E^{FF}$ is the FF field amplitude at frequency ω, $E^{SH}$ is the SH field amplitude, c is the speed of light in vacuum and $\Delta k = 2k_{FF} - k_{SH} = 2\omega[n(\omega) - n(2\omega)]/c$ is the momentum mismatch between the FF and SH waves. For normal incidence, and circularly polarized excitation of z-cut LN, the equation takes the form (see Supplementary Section 1):

$$\frac{dE^{SH}_{-\sigma}(z)}{dz} = -\frac{\omega}{n(2\omega)\,c} 2\sqrt{2}\sigma d_{22} e^{i\varphi_{GP}(\theta)}(E^{FF}_\sigma)^2 e^{i\Delta k z} \quad (2)$$

where σ marks the spin state of the circularly polarized waves (See Fig. 1b), $d_{22}$ is the relevant second-order nonlinear coefficient of the LN, and $\varphi_{GP}(\theta)$ is the geometric phase as function of the rotation angle of the LN around its z-axis.

Considering the added contribution from a thin layer of thickness l to a previously generated SH, the rotational dependance of the geometric phase offers a mechanism to control the interference of the total SH. As illustrated in Fig. 1c for a bilayer structure of TFLN, the total intensity in this case is proportional to $1 + \cos(3\theta - \Delta k l)$ (see Supplementary Section 2). Consequently, the SH can be continuously modulated from constructive interference for $\Delta k l = $ 3θ + 2πm, to destructive interference for Δkl = 3θ + π(2m + 1), opening the door for real-time continuous modulation and routing of the nonlinear signal.

This approach is now expanded to a multilayered configuration, where each TFLN layer can induce a distinct nonlinear geometric phase, as illustrated in Fig. 1d. The multilayered TFLN structure can be treated as a bulk nonlinear metamaterial with a discrete spatial variation of the effective nonlinear coefficient. For an infinite structure of successive identical layers of thickness l, where each layer adds a fixed geometric phase increment of $\Delta\varphi_{GP}$, the spatially varying nonlinear coefficient is (see Supplementary Section 3):

$$d(z) = d_{22} \sum_{m=-\infty}^{\infty} \text{sinc}\left(\frac{\Delta\varphi_{GP} + 2\pi m}{2}\right) e^{i\left(\frac{\Delta\varphi_{GP} + 2\pi m}{l}\right)z} \qquad (3)$$

The lowest order m=0 contributes a structural momentum of $k_{GP} = \Delta\varphi_{GP}/l$, that compensates for the momentum mismatch Δk when $\Delta\varphi_{GP} = \Delta k l$. Since the relative phase between the layers is $\Delta\varphi_{GP} = 3\theta$, to achieve GPM of the nonlinear interaction, every layer should be rotated at an angle of Δkl/3 relative to the previous one.

The GPM approach reveals that as the thickness of the layers is reduced, the accumulated SH approaches the ideal case of perfect PM (see Supplementary Section 3), which emphasizes the advantage of the thin-film platform. Furthermore, since the GP is decoupled from the physical constraints of the system, it can compensate for any arbitrary phase mismatch that is present in the system. Thus, the GPM approach can be applied for any combination of linear and nonlinear layers. Moreover, full control over the rotation angle removes the need for thickness precision in the design, in contrast to QPM in periodically poled crystals, where the required domain thickness is determined by the coherence length π/Δk. As such, the GPM concept is in principle a continuous generalization of the QPM method, where domain inversion poling is replaced with rotational modulation, that expands the tunability and introduces spin-selective properties. Figure 1d compares the different PM techniques, showing that multilayered thin-film crystals with thickness in the sub-micron range can obtain near-perfect PM.

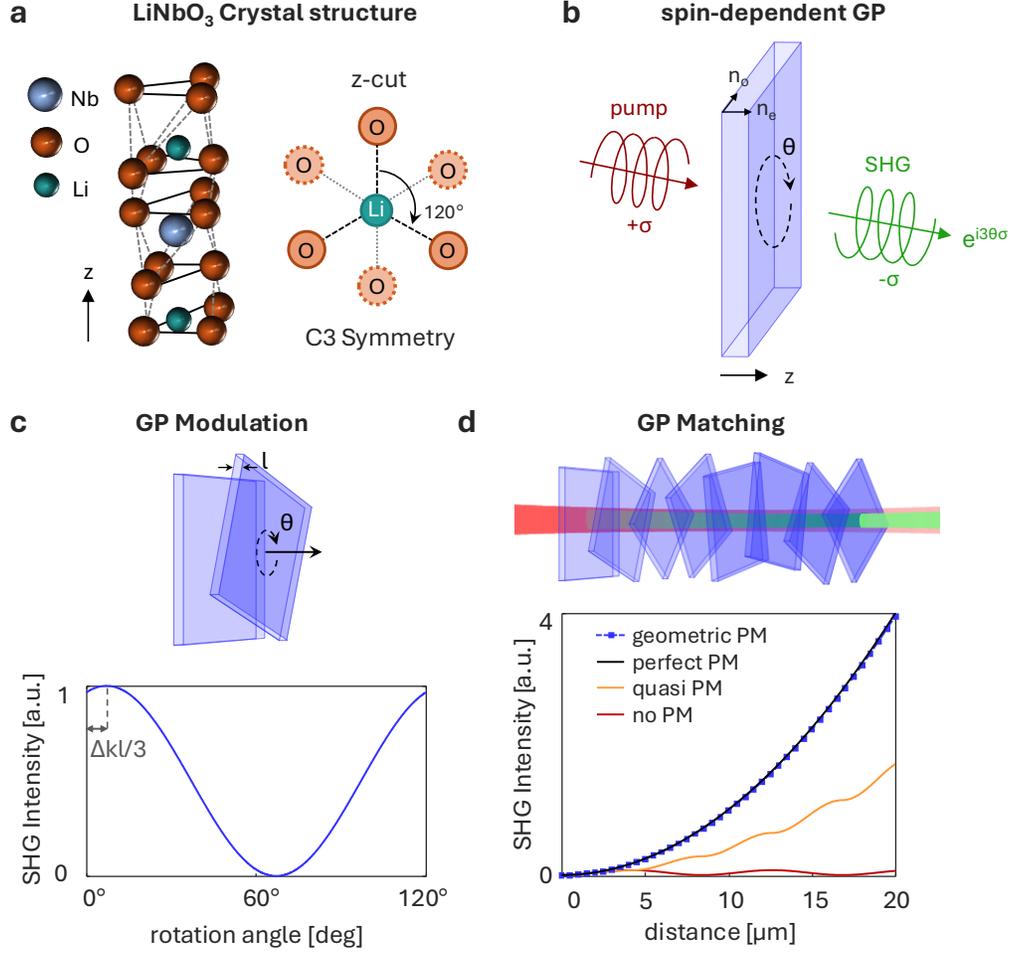

**Fig. 1: Geometric Phase Modulation and Matching of SHG in TFLN.** (a) Crystal structure of LiNbO$_3$ exhibiting C3 rotational symmetry around the z-axis. (b) Illustration of the spin-dependent nonlinear geometric phase of SHG from a z-cut TFLN. (c) Geometric phase modulation of SHG in a bilayer structure of 500 nm thick TFLN. The phase shift of the modulation is related to the layer thickness l and the momentum mismatch Δk due to the material dispersion. (d) GPM in z-cut multilayered TFLN in comparison to other PM conditions in LN. The SHG intensity as function of propagation in the z-direction is compared between the different cases; no PM (red), QPM (yellow), theoretical perfect PM (black), and GPM with 500nm thick layers (blue dotted line). In the GPM case, each consecutive thin-film layer is rotated relative to the previous layer, to phase match the nonlinear interaction.

**Experimental Results**

To validate the concept of GPM, and demonstrate real-time control over the nonlinear interaction, we perform experimental measurements on bi- and multi-layer configurations of individually mounted TFLN layers separated by air, where each layer is composed of 500 nm thick z-cut LN on top of a 0.5 mm thick silica substrate. We illuminate the structures with near-infrared light using a tunable femtosecond optical parametric oscillator (OPO) and measure the SH signal (see Fig. 2a and Methods for detailed information).

For the bilayer measurements, the first layer is mounted on a rotating stage, while the second layer remains fixed. Figure 2b shows the measured SH intensity for a fundamental wavelength of 1200 nm as function of the rotation angle θ of the first layer. In this measurement we use a mirror configuration where the LN sides are facing each other. It can be seen that by rotating the first TFLN, the transmitted SH can be modulated continuously through the geometric phase, between the maximal value and zero. Since the refractive index of the air gap can be assumed constant, this configuration introduces only the LN dispersion to the frequency conversion process, and the geometric phase compensates exclusively for the phase mismatch of the nonlinear layer. Due to the opposite orientation of the crystal's z-axis between the two nonlinear layers, the relative geometric phase between them is π − 3θ. Therefore, the minimum value of the SH intensity appears at $3θ = 2π − Δkl$. The measured value of $Δkl/3$ was $θ = 11° \pm 2°$, compared to the predicted value of $θ = 7°$ for a pair of ideal 500 nm thick TFLN layers at fundamental wavelength of 1200 nm. The small difference between the calculation and the measurement can be attributed to variations in thicknesses, minor misalignments in the experimental setup and weak Fabry-Perot propagation phase.

Another measurement was performed when both layers were placed in the same z-orientation. In this configuration, the FF and SH transmitted through the first layer accumulate an additional propagation phase when passing through the ~0.5 mm thick silica substrate. The exact optical properties and thickness of the substrate are unregulated and can vary across the substrate, practically leading to an arbitrary phase difference between the FF and SH. Nevertheless, it can be seen in Fig. 2c that the full sinusoidal modulation of the total SH can be achieved also in this case, enabling real-time geometric phase-controlled constructive or destructive interference of the SH, as well as measurement of the exact phase mismatch in the system. In addition, we measured the SH signal under excitation with both circular polarizations. Due to the antisymmetric dependence of the geometric phase on the polarization handedness, $\varphi_{GP}(-\sigma) = -\varphi_{GP}(\sigma)$, we achieve complementary modulation patterns for this case as shown in Fig. 2c, demonstrating spin-controlled modulation of the SH.

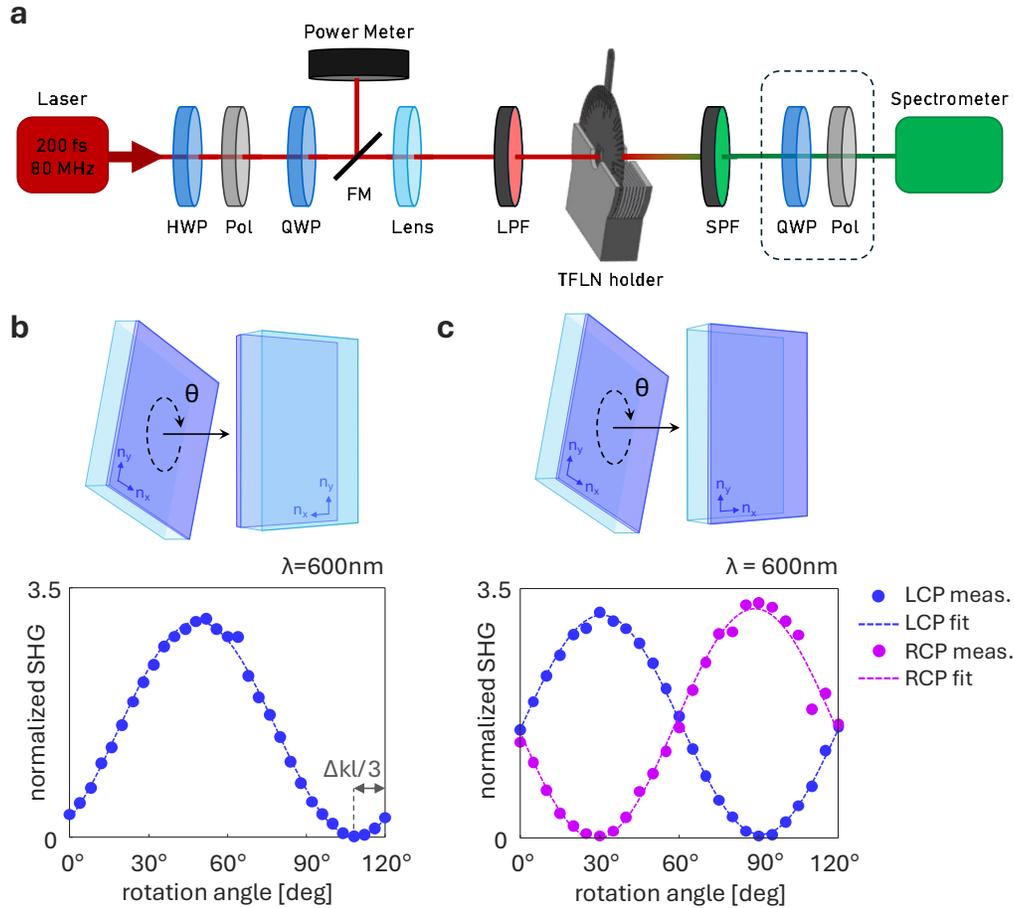

**Fig. 2: Experimental Demonstration of Spin-Controlled SHG Geometric Phase Modulation in Bilayer TFLN. (a)** Experimental setup. The samples were illuminated with polarization and power adjusted tunable femtosecond OPO, and the SH signal was measured using a spectrometer. An optional module (marked by the green dashed line) allowed analyzing the polarization state of the generated SH (HWP – half-waveplate; Pol – Linear polarizer; QWP – quarter waveplate; FM – flip mirror; LPF – longpass filter; SPF – shortpass filter. See Methods for full experimental details). **(b-c)** SHG measurements for a pump wavelength of 1200 nm from a bilayer structure of two layers separated by air, each composed of 500 nm z-cut LN film on a 0.5 mm silica substrate. In (b) the lithium niobate sides are facing each other with opposite crystal orientation, and in (c) the layers are placed with the same orientation. The SHG measurements are normalized to the single layer case with a sinusoidal fit as function of the rotation angle. In (b) the excitation is with a left circularly polarized pump (LCP), while in (c) both circular polarizations were measured, showing the spin-dependency of the geometric phase modulation.

As shown in Fig. 1d and in Eqs. (1-3), extending to multilayered TFLN platform enables full GPM and coherent spin-controlled buildup of the SH. To demonstrate this concept, we proceed with an experiment on an 8-layer configuration. The different layers were placed on a designated 3D-printed mounting system, as illustrated in Fig. 2a (see Methods and Supplementary Section 4 for the mounting system design), that allows for arbitrary and tunable control of the geometric phase in real-time by the rotation of each disk independently. We configured the multilayer to different GPM conditions and measured the SH buildup after each layer. To compare with the theoretical analysis of tightly stacked or impedance matched layers (shown in Fig. 1d), we normalize the measured results to evaluate the internal SH buildup (see Supplementary Section 5). Figure 3a shows the internal SH buildup as a function of the total number of layers, with the multilayer configuration phase-matched for a circularly polarized pump at a wavelength of 1200 nm. As can be seen, the normalized measurement results follow a nearly quadratic curve, showing excellent agreement with theory (Fig. 1d), providing a first proof of concept of GPM in this multilayered system. The absolute SH power measured from the 8-layer structure indicates an internal conversion efficiency of $\sim 9.3 \cdot 10^{-7}\%$ for peak power densities of $\sim 12$ MW/cm$^2$. A similar measurement performed using an optical parametric amplifier yielded an internal conversion efficiency of $\sim 4.7 \cdot 10^{-5}\%$ from a single TFLN layer for peak power densities of $\sim 49$ GW/cm$^2$. For the 8-layer structure, this value scales to internal conversion efficiency of $\sim 2.7 \cdot 10^{-3}\%$, and is expected to continue to grow quadratically with the number of layers.

To demonstrate the selectivity of the GPM condition, Fig. 3a includes measurements from the 8-layer configuration at a detuned pump wavelength of 1150 nm, while maintaining the same rotation angles that were configured to phase-match the system at wavelength of 1200 nm. This is a non-phase-matched case in which the buildup of the SH signal is suppressed due to the additional phase mismatch introduced by the substrate, and typical oscillations are observed in the measurements. To further illustrate the tunability of the GPM method, we switch between the wavelengths and reconfigure the system in real-time to be phase-matched at pump wavelength of 1150 nm, as shown in Fig. 3b. As expected, the normalized results are very close to the perfect PM case when pumped with 1150 nm, while in this case the 1200 pump is non-phase-matched and does not show a coherent buildup of the SH.

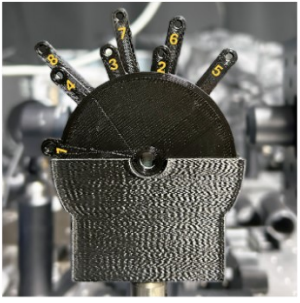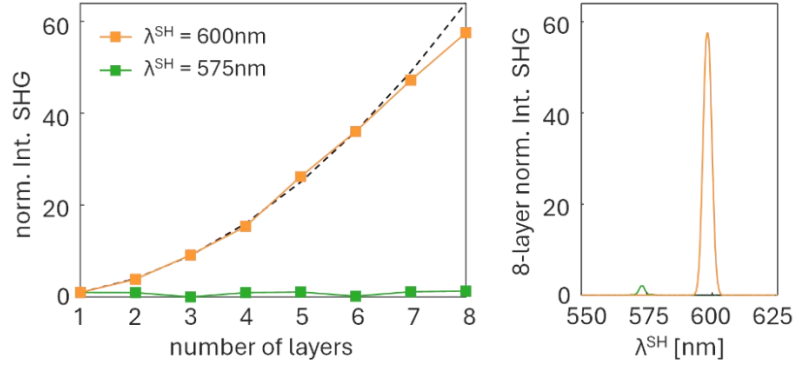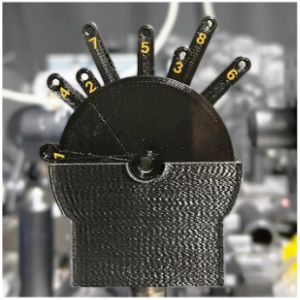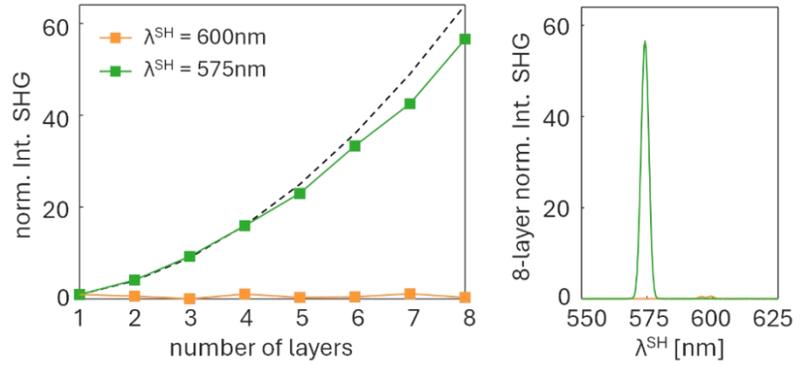

**Fig. 3: Experimental Demonstration of Real-Time Frequency-Dependent GPM in an 8-Layer TFLN.** **(a-b)** Experimental demonstration of the frequency-dependent GPM in an 8-layer configuration, using an LCP pump. The configuration in (a) was designed to phase-match the SHG for fundamental wavelength of 1200 nm, and in (b) for 1150 nm. From left to right: Image of the experimental mounting system from the optical axis point-of-view, the number of each layer is indicated on the arms of the rotation disks that hold the LN sample layers; The measured internal SHG normalized to the single layer case, as function of the number of layers (the dashed line represents the quadratic dependence on the number of layers); The spectral measurement of the 8-layer internal SHG normalized to the single layer peak value.

To measure the structural spin-selectivity of the GPM multilayer, we perform polarization tomography of the generated SH at the output of the multilayered system when excited with linearly polarized light. As illustrated in Fig. 4a, we expect only one spin-state to be phase matched, and therefore to observe a quadratic increase of a single spin state over the other, leading to a pure SH spin state for a sufficient number of layers. Figure 4b shows the expected gradual evolution across the layers from linear to an almost circular polarization state, based on finite-element simulations (see Methods). The experimental results, measured for a pump wavelength of 1200 nm, are shown in Fig. 4c, which displays the extracted polarization state on the Poincare sphere and by the polarization ellipse. To calculate the polarization state, the SHG intensity was sampled using a quarter-waveplate and a polarizer (See Fig. 2a), across a 180° rotation range of the quarter-waveplate (see Supplementary Section 6). The SH polarization at the output of the 8-layer structure is nearly circularly polarized, (ellipticity of ~0.7) and in excellent agreement with the simulation results shown in Fig. 4b. Setting the rotation angles of the layers for GPM of the opposite spin state yielded similar results for linear excitation, but for the opposite SH circular polarization.

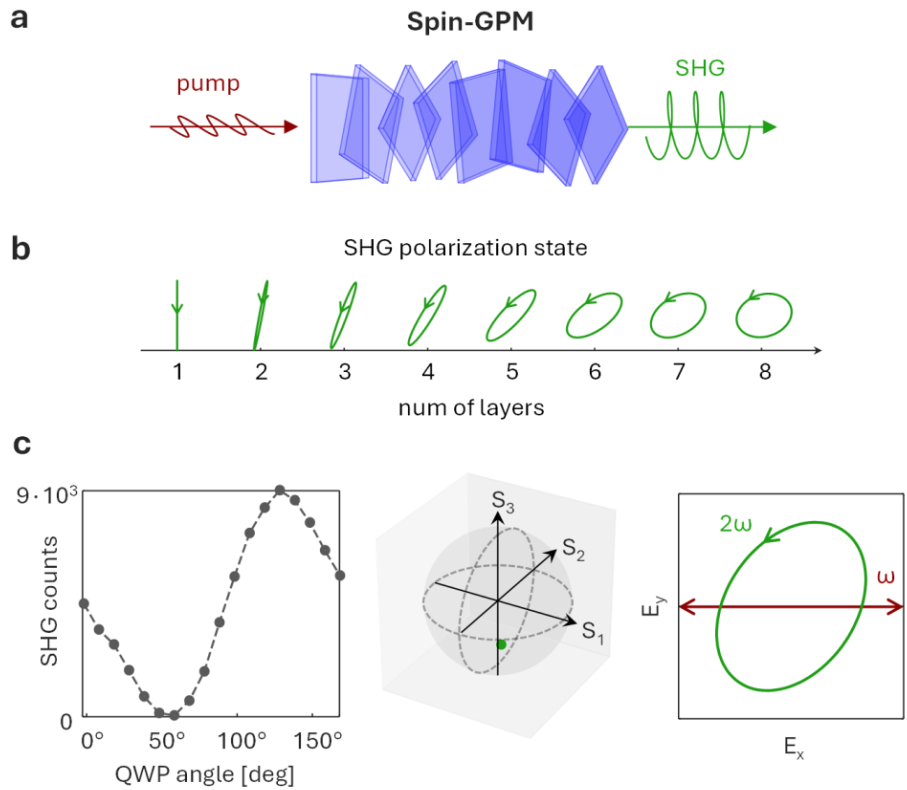

**Fig. 4: Spin-Selective GPM in an 8-Layer TFLN. (a)** Illustration of spin-selective GPM. The layers are rotated to match the SHG interaction of only one spin state, making this SH spin state dominant under linear pump excitation. **(b)** Evolution of the SHG polarization state as function of the number of layers, for the configuration illustrated in (a), based on finite-element simulations. **(c)** Experimental demonstration of spin-GPM for 8 layers, measuring the SHG polarization state for linear pump excitation. From left to right: SHG intensity as function of the angle of the QWP close to the spectrometer, the calculated point on the Poincare sphere, and the extracted polarization ellipse.

**Conclusions**

We have theoretically and experimentally demonstrated the concept of reconfigurable GPM in bi- and multi-layer TFLN as a generalized PM approach to construct nearly perfectly phase-matched nonlinear crystals, highlighting its remarkable capability to manipulate the nonlinear generation of light in new ways and in real-time. The bilayer configuration enabled continuous analog and spin-controlled binary modulation of SHG with complete dynamic contrast, which was experimentally demonstrated using 500 nm-thick TFLN. This advancement enables potential applications in signal processing, multiplexing, and routing of nonlinear optical signals. We then leveraged this approach to realize reconfigurable and nearly perfect PM in multilayer TFLN structures. Specifically, by employing an eight-layer TFLN configuration, we experimentally attained up to 57-fold growth in internal SHG, growing quadratically with the number of layers, and approaching the theoretical factor of 64-fold. The external SHG output can approach these values as well by minimizing reflections in the system. This could be done by incorporating anti-reflection coatings, or by direct bonding between the nonlinear thin-films to create a multilayered nonlinear metamaterial. We also demonstrated how the dispersion-independent nature of the geometric phase enables broad spectral tuning within the same multilayer TFLN platform, in contrast to QPM techniques. Furthermore, we demonstrated the intrinsic structural spin effect on the nonlinear interaction by performing polarization tomography and realizing nonlinear polarization conversion in the 8-layer TFLN from linear to circular. These demonstrations establish a versatile methodology that can readily extend beyond SHG processes to general wave mixing applications, facilitating for example controlled generation of exotic frequencies in spectral regions such as far-infrared or terahertz (THz), as well as the generation of single and entangled photons. Finally, although our current demonstrations utilize unstructured TFLN layers, future incorporation of nano-structured LN within the multilayers could significantly expand the accessible functionalities via full three-dimensional structural GPM and nonlinear manipulation of light, paving the way for new types of three-dimensional nonlinear photonic crystals[51,52] and nonlinear metamaterials. Collectively, our results introduce a highly scalable, robust, and tunable platform for nonlinear optics, opening numerous exciting avenues for advanced adaptable nonlinear photonic control and tailored light generation.

**Methods:**

*Sample preparation and 3D-printed mounting system design*

The samples were prepared by cutting 6.25 mm x 6.25 mm squares from a commercially available wafer of 500 nm z-cut lithium niobate on top of 0.5 mm silica substrate, using a high-precision dicer. The 3D-printed mounting system comprised of circular rotation disks, each featuring a hole in their center for mounting the sample, and a base holder with slots in which the rotation disks were inserted. The rotation disks featured arms of increasing length for easy rotation and identification of the layer number. The rotation disks were designed with an outer diameter of 60mm and a thickness of 1mm. When the rotation disks were placed inside the slots, the spacing between adjacent samples was approximately 1.5 mm.

*Experimental setup for SHG measurements*

The exciting laser was a tunable femtosecond OPO (Coherent Chameleon; ~200 fs pulse width, 80 MHz repetition rate). The laser power was controlled using a half-waveplate (HWP) followed by a linear polarizer (pol), while maintaining a fixed linear polarization state. A quarter-waveplate (QWP) was then used to control the pump polarization state, allowing to switch between linear and circular polarizations. A flip mirror allowed to redirect the pump to a power meter. The pump beam was weakly focused onto the TFLN samples using a lens (L) with focal length f = 300 mm. A long-pass filter (LPF) was used to filter out parasitic SH signals. The TFLN layers were placed on the reconfigurable 3D-printed mounting system, which could accommodate up to 8 separately rotatable TFLN layers. The mounting system was followed by a short-pass filter (SPF), to filter out the pump light. After the SPF, there was an optional module including a QWP followed by a polarizer, to analyze the polarization state of the generated SH. Finally, the SH signal was directed into an imaging spectrometer (Andor Shamrock 303). All SH measurements taken by the spectrometer were normalized to the square of the pump power measured by the power meter, to compensate for fluctuations in the laser power.

*Finite-element Simulations*

We performed finite-element simulations using COMSOL Multiphysics to analyze the SHG polarization state from the multilayer TFLN. The frequency-domain nonlinear simulation was computed in two steps. In the first step, a FF plane wave was generated at the input port, and the FF field was computed across the structure. It was then used to calculate the nonlinear polarization at the SH frequency, as a source for the second step. The nonlinear polarization was calculated according to the second-order nonlinear tensor, rotated in each layer at the corresponding PM angle. The SH polarization ellipse was derived from the SH field, recorded at the output port.

**References:**

1. Boes, A. *et al.* Lithium niobate photonics: Unlocking the electromagnetic spectrum. *Science* **379**, (2023).
2. Ghimire, S. & Reis, D. A. High-harmonic generation from solids. *Nat. Phys.* **15**, 10–16 (2019).
3. Hebling, J., Yeh, K. Lo, Hoffmann, M. C. & Nelson, K. A. High-power THz generation, THz nonlinear optics, and THz nonlinear spectroscopy. *IEEE J. Sel. Top. Quantum Electron.* **14**, 345–353 (2008).
4. Couteau, C. Spontaneous parametric down-conversion. *Contemp. Phys.* **59**, 291–304 (2018).
5. Santiago-Cruz, T. *et al.* Resonant metasurfaces for generating complex quantum states. *Science* **377**, 991–995 (2022).
6. Saravi, S., Pertsch, T. & Setzpfandt, F. Lithium Niobate on Insulator: An Emerging Platform for Integrated Quantum Photonics. *Adv. Opt. Mater.* **9**, 2100789 (2021).
7. Pupeza, I., Zhang, C., Högner, M. & Ye, J. Extreme-ultraviolet frequency combs for precision metrology and attosecond science. *Nat. Photonics* **15**, 175–186 (2021).
8. Wu, T. H. *et al.* Visible-to-ultraviolet frequency comb generation in lithium niobate nanophotonic waveguides. *Nat. Photonics* **18**, 218–223 (2024).
9. Yu, M. *et al.* Integrated femtosecond pulse generator on thin-film lithium niobate. *Nature* **612**, 252–258 (2022).
10. Armstrong, J. A., Bloembergen, N., Ducuing, J. & Pershan, P. S. Interactions between light waves in a nonlinear dielectric. *Phys. Rev.* **127**, 1918–1939 (1962).
11. Maker, P. D., Terhune, R. W., Nisenoff, M. & Savage, C. M. Effects of dispersion and focusing on the production of optical harmonics. *Phys. Rev. Lett.* **8**, 21–22 (1962).
12. Giordmaine, J. A. Mixing of light beams in crystals. *Phys. Rev. Lett.* **8**, 19–20 (1962).
13. Boyd, G. D., Miller, R. C., Nassau, K., Bond, W. L. & Savage, A. LiNbO3: An efficient phase matchable nonlinear optical material. *Appl. Phys. Lett.* **5**, 234–236 (1964).
14. Li, G., Zhang, S. & Zentgraf, T. Nonlinear photonic metasurfaces. *Nat. Rev. Mater.* **2**, 1–14 (2017).
15. Krasnok, A., Tymchenko, M. & Alù, A. Nonlinear metasurfaces: a paradigm shift in nonlinear optics. *Mater. Today* **21**, 8–21 (2018).
16. Kadic, M., Milton, G. W., van Hecke, M. & Wegener, M. 3D metamaterials. *Nat. Rev. Phys.* **1**, 198–210 (2019).

17. Segal, N., Keren-Zur, S., Hendler, N. & Ellenbogen, T. Controlling light with metamaterial-based nonlinear photonic crystals. *Nat. Photonics* **9**, 180–184 (2015).

18. Li, G. *et al.* Continuous control of the nonlinearity phase for harmonic generations. *Nat. Mater.* **14**, 607–612 (2015).

19. Gigli, C. *et al.* Tensorial phase control in nonlinear meta-optics. *Optica* **8**, 269 (2021).

20. Marino, G. *et al.* Harmonic generation with multi-layer dielectric metasurfaces. *Nanophotonics* **10**, 1837–1843 (2021).

21. Stolt, T. *et al.* Backward Phase-Matched Second-Harmonic Generation from Stacked Metasurfaces. *Phys. Rev. Lett.* **126**, 033901 (2021).

22. Ben-Haim, D. & Ellenbogen, T. Optical Anomalies due to Volume Collective Modes of Plasmonic Metamaterials. *Laser Photonics Rev.* **17**, 2200671 (2023).

23. Tang, Y. *et al.* Quasi-phase-matching enabled by van der Waals stacking. *Nat. Commun.* **15**, 9979 (2024).

24. Trovatello, C. *et al.* Quasi-phase-matched up- and down-conversion in periodically poled layered semiconductors. *Nat. Photonics* **19**, 291–299 (2025).

25. Kim, B. *et al.* Three-dimensional nonlinear optical materials from twisted two-dimensional van der Waals interfaces. *Nat. Photonics* **18**, 91–98 (2024).

26. Hong, H. *et al.* Twist Phase Matching in Two-Dimensional Materials. *Phys. Rev. Lett.* **131**, 233801 (2023).

27. Berry, M. V. Quantal phase factors accompanying adiabatic changes. *Proc. R. Soc. London. A. Math. Phys. Sci.* **392**, 45–57 (1984).

28. Pancharatnam, S. Generalized theory of interference, and its applications - Part I. Coherent pencils. *Proc. Indian Acad. Sci. - Sect. A* **44**, 247–262 (1956).

29. Hu, Z. & Li, G. Nonlinear geometric phase in optics: Fundamentals and applications. *Appl. Phys. Lett.* **126**, 100502 (2025).

30. Lin, D., Fan, P., Hasman, E. & Brongersma, M. L. Dielectric gradient metasurface optical elements. *Science* **345**, 298–302 (2014).

31. Balthasar Mueller, J. P., Rubin, N. A., Devlin, R. C., Groever, B. & Capasso, F. Metasurface Polarization Optics: Independent Phase Control of Arbitrary Orthogonal States of Polarization. *Phys. Rev. Lett.* **118**, 113901 (2017).

32. Dorrah, A. H., Tamagnone, M., Rubin, N. A., Zaidi, A. & Capasso, F. Introducing Berry phase gradients along the optical path via propagation-dependent polarization transformations. *Nanophotonics* **11**, 713–725 (2022).

33. Tymchenko, M. *et al.* Gradient Nonlinear Pancharatnam-Berry Metasurfaces. *Phys. Rev. Lett.* **115**, 207403 (2015).

34. Li, G. *et al.* Continuous control of the nonlinearity phase for harmonic generations. *Nat. Mater.* **14**, 607–612 (2015).

35. Lu, L., Joannopoulos, J. D. & Soljačić, M. Topological photonics. *Nat. Photonics* **8**, 821–829 (2014).

36. Karnieli, A., Li, Y. & Arie, A. The geometric phase in nonlinear frequency conversion.

*Front. Phys.* **17**, 1–31 (2022).

37. Tal, M., Haim, D. Ben & Ellenbogen, T. Geometric phase opens new frontiers in nonlinear frequency conversion of light. *Front. Phys.* **17**, 1–5 (2022).

38. Cohen, E. *et al.* Geometric phase from Aharonov–Bohm to Pancharatnam–Berry and beyond. *Nat. Rev. Phys.* **1**, 437–449 (2019).

39. Myers, L. E. & Bosenberg, W. R. Periodically poled lithium niobate and quasi-phase-matched optical parametric oscillators. *IEEE J. Quantum Electron.* **33**, 1663–1672 (1997).

40. Honardoost, A., Abdelsalam, K. & Fathpour, S. Rejuvenating a Versatile Photonic Material: Thin-Film Lithium Niobate. *Laser Photonics Rev.* **14**, 2000088 (2020).

41. Zhu, D. *et al.* Integrated photonics on thin-film lithium niobate. *Adv. Opt. Photonics* **13**, 242 (2021).

42. Vazimali, M. G. & Fathpour, S. Nonlinear integrated photonics in thin-film lithium niobate. *Adv. Nonlinear Photonics* **4**, 215–246 (2023).

43. Pohl, D. *et al.* An integrated broadband spectrometer on thin-film lithium niobate. *Nat. Photonics* **14**, 24–29 (2020).

44. Fedotova, A. *et al.* Lithium Niobate Meta-Optics. *ACS Photonics* **9**, 3745–3763 (2022).

45. Li, G. *et al.* Nonlinear Metasurface for Simultaneous Control of Spin and Orbital Angular Momentum in Second Harmonic Generation. *Nano Lett.* **17**, 7974–7979 (2017).

46. Ye, W. *et al.* Spin and wavelength multiplexed nonlinear metasurface holography. *Nat. Commun.* **7**, 11930 (2016).

47. Walter, F., Li, G., Meier, C., Zhang, S. & Zentgraf, T. Ultrathin Nonlinear Metasurface for Optical Image Encoding. *Nano Lett.* **17**, 3171–3175 (2017).

48. McDonnell, C., Deng, J., Sideris, S., Ellenbogen, T. & Li, G. Functional THz emitters based on Pancharatnam-Berry phase nonlinear metasurfaces. *Nat. Commun.* **12**, 30 (2021).

49. Zdagkas, A. *et al.* Observation of toroidal pulses of light. *Nat. Photonics* **16**, 523–528 (2022).

50. Boyd, R. W. *Nonlinear Optics*. *Nonlinear Optics* (Academic Press, 2020). doi:10.1016/C2015-0-05510-1

51. Xu, T. *et al.* Three-dimensional nonlinear photonic crystal in ferroelectric barium calcium titanate. *Nat. Photonics* **12**, 591–595 (2018).

52. Wei, D. *et al.* Experimental demonstration of a three-dimensional lithium niobate nonlinear photonic crystal. *Nat. Photonics* **12**, 596–600 (2018).

**Acknowledgments:**

This research was supported by the European Research Council (ERC 3D NOAM 101044797). D.B.H acknowledges the doctoral scholarship given by the Council for Higher Education (CHE) of Israel. M.T. acknowledges the support of the Milner Foundation fellowship for PhD students. We thank Rachel Shmuel from the Tel-Aviv Unviersity Engineering Workshop for fabrication of the multilayered mounting system.

**Additional information:**

Correspondence and requests for materials should be addressed to T. Ellenbogen and D. Ben-Haim.

**Competing interests statement:**

The authors declare that they have no competing financial interests.

# Supplementary Information

## Section 1: TFLN Nonlinear coefficients

The nonlinear tensor (in contracted notation) of a 3m class crystal has the following form[1]:

$$d_{il} = \begin{pmatrix} 0 & 0 & 0 & 0 & d_{31} & -d_{22} \\ -d_{22} & d_{22} & 0 & d_{31} & 0 & 0 \\ d_{31} & d_{31} & d_{33} & 0 & 0 & 0 \end{pmatrix} \quad (S1)$$

where the nonlinear coefficients for lithium niobate[2] are $d_{22} = 2.1 \text{ pm/V}$, $d_{31} = 4.6 \text{ pm/V}$, $d_{33} = 25.2 \text{ pm/V}$. For normal incident pump, propagating along the z-axis, the nonlinear polarization field in the crystal coordinate system is:

$$P_x^{NL} = -4d_{22}(E_x E_y), \quad P_y^{NL} = -2d_{22}(E_x^2 - E_y^2), \quad P_z^{NL} = 2d_{31}(E_x^2 + E_y^2) \quad (S2)$$

If we add rotation around the z-axis at angle θ, and excite with a circularly polarized pump with spin state σ, we get:

$$P_{-\sigma}^{NL} = 2\sqrt{2}\, d_{22} i\sigma e^{i\sigma 3\theta}(E_\sigma^{FF})^2 \quad (S3)$$

The governing equation for SHG (see eq. 2 in the manuscript) with a general spatially varying nonlinear coefficient d(z) is:

$$\frac{dE_{-\sigma}^{SH}(z)}{dz} = -\frac{\omega}{n(2\omega)c} 2\sqrt{2}\sigma\, d(z)(E_\sigma^{FF})^2 e^{i\Delta kz} \quad (S4)$$

where in the case of the multilayered TFLN $d(z) = d_{22} e^{i\sigma 3\theta(z)}$. The solution is achieved by integrating the governing equation across the structure. By applying integration over $(-\infty, \infty)$ we get the following solution:

$$E_{-\sigma}^{SH}(z) = \underbrace{-\frac{\omega}{n(2\omega)c} 2\sqrt{2}\sigma\, (E_\sigma^{FF})^2}_{A} \underbrace{\int_{-\infty}^{\infty} d(z) e^{i\Delta kz} dz}_{FT_{\Delta k}\{d(z)\}} \quad (S5)$$

where d(z) is defined within $z \in [0, L]$, and L is the structure's length.

## Section 2: GP Modulation in Bilayer configuration

The bilayer structure with layer's thickness l, and a relative geometric phase $\varphi_{GP}$ between the second and the first layer, is described by:

$$d(z) = d_{22}\left[\text{rect}\left(\frac{z}{l}\right) + e^{i\varphi_{GP}}\text{rect}\left(\frac{z-l}{l}\right)\right] \quad (S6)$$

and thus by applying eq. S6 to eq. S5 the transmitted SHG is:

$$E_{-\sigma}^{SH} = A(E_\sigma^{FF})^2 d_{22}\, 2\,\text{sinc}\left(\frac{\Delta kl}{2}\right)\cos\left(\frac{\varphi_{GP}-\Delta kl}{2}\right)e^{\frac{i(\varphi_{GP}-\Delta kl)}{2}} \quad (S7)$$

Therefore, the second harmonic intensity goes as $I_{-\sigma}^{SH} \propto 1 + \cos(\varphi_{GP} - \Delta kl)$.

## Section 3: GP Matching in multilayer configuration

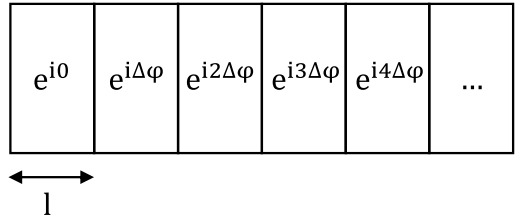

**Fig. S1: Discrete variation of the nonlinear coefficient across a multilayer structure.** Each layer of thickness l contributes a nonlinear phase increment of $\Delta\varphi$.

The discrete spatial variation of the nonlinear phase across an infinite multilayer structure, as shown in Fig. S1, is described as follows:

$$d(z) = d_{22}\sum_{m=-\infty}^{\infty} e^{im\Delta\varphi}\text{rect}\left(\frac{z-ml}{l}\right) = d_{22}\text{rect}\left(\frac{z}{l}\right) * \left[e^{\frac{iz\Delta\varphi}{l}}\sum_{m=-\infty}^{\infty}\delta(z-ml)\right] \quad (S8)$$

by applying a Fourier transform and then its inverse, we get the harmonic decomposition of d(z):

$$FT_{\Delta k}\{d(z)\} = d_{22}\,\text{sinc}\left(\frac{\Delta kl}{2}\right)\left[\sum_{m=-\infty}^{\infty}\delta\left(\Delta k - \frac{\Delta\varphi + 2\pi m}{l}\right)\right] = \quad (S9)$$

$$= d_{22}\sum_{m=-\infty}^{\infty}\text{sinc}\left(\frac{\Delta\varphi + 2\pi m}{2}\right)\delta\left(\Delta k - \frac{\Delta\varphi + 2\pi m}{l}\right),$$

$$d(z) = FT^{-1}\{FT_{\Delta k}\{d(z)\}\} = d_{22}\sum_{m=-\infty}^{\infty}\text{sinc}\left(\frac{\Delta\varphi + 2\pi m}{2}\right)e^{i\left(\frac{\Delta\varphi + 2\pi m}{l}\right)z}$$

Assuming a finite structure of N layers and applying it to eq. S5, we get the transmitted SHG:

$$E^{SH}_{-\sigma} = A\,(E^{FF}_\sigma)^2\,d_{22}\cdot$$
$$\sum_{m=-\infty}^{\infty} \operatorname{sinc}\left(\frac{\Delta\varphi + 2\pi m}{2}\right) Nl\, \operatorname{sinc}\left(\frac{N}{2}(\Delta kl - \Delta\varphi - 2\pi m)\right) e^{-\frac{i}{2}(N+1)(\Delta kl - \Delta\varphi - 2\pi m)} \quad (S10)$$

Therefore, the dominant harmonic of the multilayer structure contributes a momentum of $\Delta\varphi/l$ that can compensate the momentum mismatch for $\Delta\varphi = \Delta kl$. It can be seen from eq. S10 that the SH is maximal at this condition.

## *Section 4: Image of the 3D-printed mounting system*

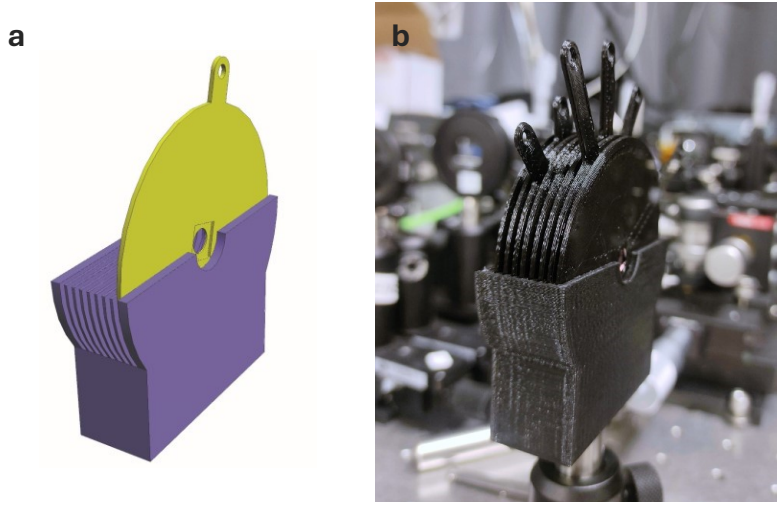

**Fig. S2: The 3D-printed mounting system. (a)** Illustration of the design 3D model. **(b)** The fabricated system in the experimental setup.

## *Section 5: Internal SHG evaluation*

In order to evaluate the internal SHG efficiency in the experimental setup of the multilayer TFLN, we normalize the measured SH with respect to the transmission of the FF and the SH waves passing through the layers. The contribution of each TFLN layer to the transmitted SH depends on both the FF and SH transmission through the layers, as described in Fig. S3. In practice, the rotation disks in the 3D-printed holder were not precisely aligned to maintain parallel interfaces of the mounted sample. This resulted in negligible cavity-related effects in the thick silica substrates and air gaps between the TFLN layers, and dominant contribution of the TFLN layers. Therefore, for our evaluation we only account for the inner reflection within the TFLN and calculate its transmission based on a Fabry Perot cavity. The coefficient $\eta_{NL}$ describes the generated SH wave from each TFLN with respect to the FF wave at its input interface. Overall, the external SH wave at the output of the multilayer structure is the sum of contributions from all layers, as follows:

$$E_N^{SH} = e^{i\phi_0}\eta_{NL}(t_s(\omega)E^{FF})^2 \sum_{n=1}^{N}[t_s(\omega)t_{LN}(\omega)]^{2(n-1)}[t_s(2\omega)t_{LN}(2\omega)]^{(N-n)}e^{in\Delta\phi} \quad (S11)$$

Where $t_s$ and $t_{LN}$ are the transmission coefficients defined in Fig. S3, $\phi_0$ is the common phase factor, and $\Delta\phi$ is the overall phase mismatch accumulated in one sample. Using the GPM approach, we compensate for the phase mismatch in both the TFLN and the silica. Thus, in the phase-matched case the external SH intensity is:

$$I_{ext}^{SH} = \eta_{NL}^2 |t_s(\omega)E^{FF}|^4 \left[\sum_{n=1}^{N}|t_s(\omega)t_{LN}(\omega)|^{2(n-1)}|t_s(2\omega)t_{LN}(2\omega)|^{(N-n)}\right]^2 \quad (S12)$$

In order to evaluate the internal SH intensity from the measurements, which is:

$$I_{int}^{SH} = \eta_{NL}^2 N^2 |E^{FF}|^4 \quad (S13)$$

We divide the measured intensity as follows:

$$I_{int}^{SH} = I_{meas}^{SH} / \left[\frac{1}{N^2}|t_s(\omega)|^4 \left[\sum_{n=1}^{N}|t_s(\omega)t_{LN}(\omega)|^{2(n-1)}|t_s(2\omega)t_{LN}(2\omega)|^{(N-n)}\right]^2\right] \quad (S14)$$

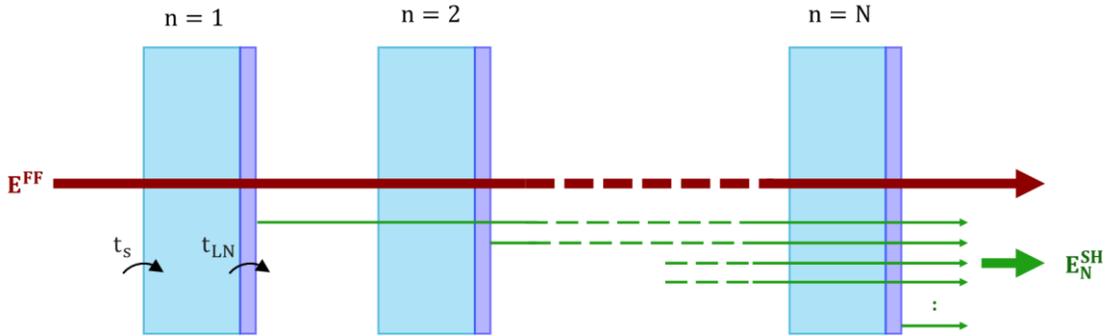

**Fig. S3: Illustration of the FF and SH waves transmission through the multilayered TFLN.** $E^{FF}$ is the input FF wave, $E_N^{SH}$ is the output SH wave, $t_s$ is the transmission into the silica, and $t_{LN}$ is the transmission through the TFLN layer.

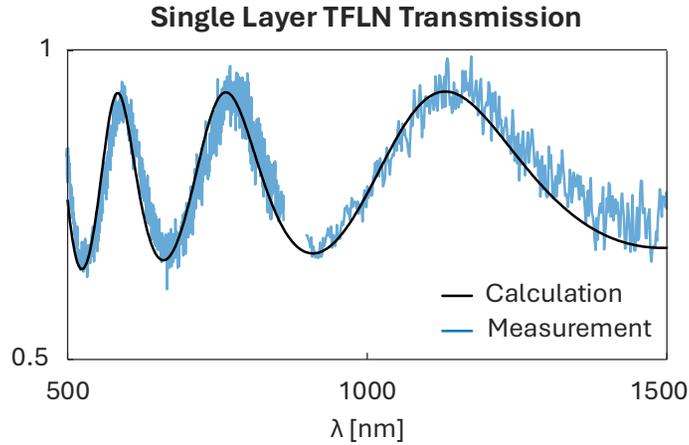

**Fig. S4: Single layer TFLN transmission spectrum.** Comparison between the calculated (black) and the measured (blue) linear power transmission from the TFLN sample, comprising of a 500 nm thick z-cut TFLN layer on top of a 0.5 mm thick silica substrate. Two separate measurements were made: for wavelengths of 500-860 nm using Andor Shamrock 303, and for 900-1500 nm using Ocean Optics NIRQuest.

In Fig. S5 we show the measured external SH normalized to the single layer case, when optimizing the system to GPM of fundamental wave of 1200nm (Fig. S5a), and 1150nm (Fig. S5b). It can be seen that the external SH grows with the number of layers, however the FH and SH reflections through the system prevent the quadratic growth with respect to the number of layers. Excluding these reflections leads to the results shown in the main text in Fig. 3a and 3b, agreeing with theory and demonstrating the internal quadratic growth with the number of layers. This can be achieved in the lab by tightly stacking the TFLN layers, or alternatively by using proper anti-reflect coatings.

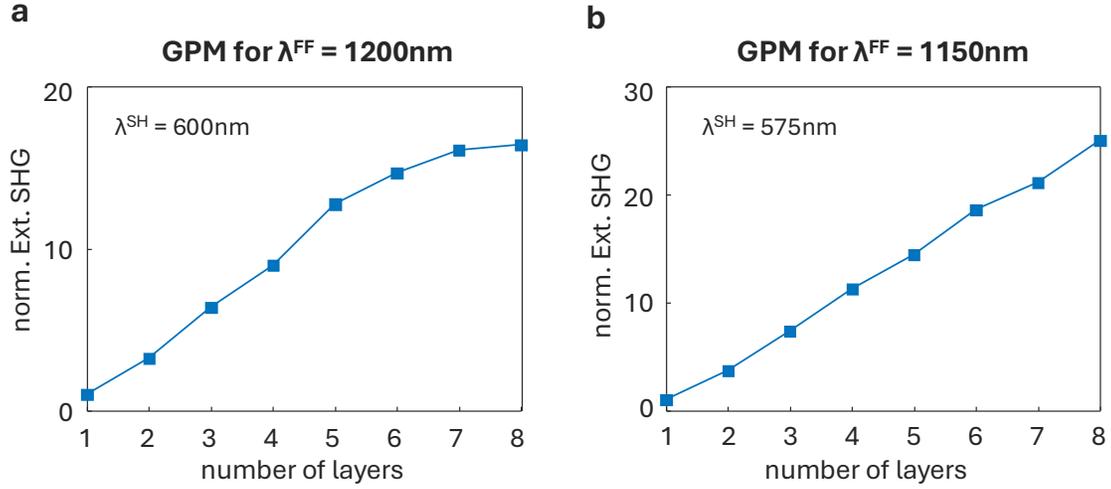

**Fig. S5: Experimental measurements of the external SHG.** The measured external SHG normalized to the single layer case, as function of the number of layers. (a) in case of PM the SHG for fundamental wavelength of 1200 nm, and (b) for 1150 nm.

## *Section 6: Measurement of the SHG polarization state*

For polarized light that passes through a rotating quarter waveplate (QWP) followed by a linear polarized, the intensity goes as[3]:

$$I(\theta) = \frac{1}{2}(E + B\sin 2\theta + C\cos 4\theta + D\sin 4\theta) \tag{S15}$$

where $\theta$ is the angle between the slow axis and the polarizer's transmission axis, and the coefficients B,C,D,E are related to the Stokes parameters $(S_0, S_1, S_2, S_3)$ by:

$$S_0 = E - C, S_1 = 2C, S_2 = 2D, S_3 = B \tag{S16}$$

We can represent eq. S15 by its Fourier series, as follows:

$$I(\theta) = A_0 + A_2 e^{2\theta} + A_{-2} e^{-2\theta} + A_4 e^{4\theta} + A_{-4} e^{-4\theta} \tag{S17}$$

Such that the Fourier coefficients $A_0, A_2, A_{-2}, A_4, A_{-4}$ equal to:

$$E = 2A_0, B = 2i \cdot (A_2 - A_{-2}), C = 2 \cdot (A_4 + A_{-4}), D = 2i \cdot (A_4 - A_{-4}) \tag{S18}$$

We can extract the relation between the Fourier coefficients to the Stokes parameters from eq. S16. To characterize the polarization of the SHG, we perform measurements for several angles of the QWP across the 180° period of the measured intensity (as seen from eq. S15). The QWP is placed at the output of the experimental setup (see Fig. 2a in the manuscript). The Fourier coefficients are calculated from the measurements, from which the Stokes parameters are derived.

**References:**

1. Boyd, R. W. *Nonlinear Optics*. (Academic Press, 2003).
2. Nikogosyan, D. N. *Nonlinear optical crystals: a complete survey*. (Springer, 2005).
3. Flueraru, C., Latoui, S., Besse, J. & Legendre, P. Error analysis of a rotating quarter-wave plate stokes' polarimeter. *IEEE Trans. Instrum. Meas.* **57**, 731–735 (2008).